\documentclass[letterpaper, 10 pt, conference]{ieeeconf}  

\IEEEoverridecommandlockouts                              

\usepackage{cite}
\usepackage{array}
\usepackage{subcaption}
\usepackage{color}
\usepackage{amsmath,amssymb}
\usepackage{algorithmicx}
\usepackage{mathrsfs}

\usepackage{ifthen}
\newboolean{showcomments}
\setboolean{showcomments}{true}
\ifx\pdfoutput\undefined
\usepackage{graphicx}
\else
\usepackage[pdftex]{graphicx}
\usepackage{epstopdf}
\fi

\newcommand{\souvik}[1]{  \ifthenelse{\boolean{showcomments}}
{\textcolor{blue}{(Souvik says:  #1)}}{}}

\title{\LARGE \bf
Exploring the Impact of Wind Penetration on Power System Equilibrium Using a Numerical Continuation Approach
}

\author{ Souvik Chandra, Dhagash Mehta and Aranya Chakrabortty
\thanks{S. Chandra and A. Chakrabortty are with Electrical and Computer Engineering Dept., North Carolina State University, Raleigh, NC, USA 27695. {\tt schandr7@ncsu.edu, aranya.chakrabortty@ncsu.edu}\newline
D. Mehta is with the Dept. of Applied and Computational Mathematics and Statistics, The University of Notre Dame, Notre Dame, IN, USA 46556. {\tt dmehta@nd.edu}
\newline
Support from the NSF ECCS grant numbers 1230848 and DARPA Young Faculty Award is gratefully acknowledged.}}

\begin{document}
\maketitle
\thispagestyle{empty}
\pagestyle{empty}

\begin{abstract}

In this paper we investigate how the equilibrium characteristics of conventional power systems may  change with an increase in wind penetration. We first derive a differential-algebraic model of a  power system network consisting of synchronous generators, loads and a wind power plant modeled by a wind turbine and a doubly-fed induction generator (DFIG). The models of these three components  are coupled via nonlinear power flow equations. In contrast to the traditional approach for solving the power flows via iterative methods that often lead to only local solutions, we apply a recently developed parameter-homotopy based numerical continuation algorithm to compute all possible solutions.  The method solves the power flow equations over multiple values of the wind penetration level with far less computational effort instead of solving them at each value individually.  We observe that depending on the penetration limit and the setpoint value for the magnitude of the wind bus voltage, the system may exhibit several undesired or even unstable equilibria. We illustrate these results through a detailed simulation of a 5-machine power system model with wind injection, and highlight how the solutions may be helpful for small-signal stability assessment. 


\end{abstract}

\section{Introduction}
\label{sec:intro}
With the increase in renewable generation such as wind and solar power as well as the intrusion of smart loads such as plug-in hybrid vehicles and smart buildings, the operational characteristics of power systems are gradually becoming more complex \cite{bose}. Since the dynamic models of these renewable sources and loads are inherently nonlinear, and very often stochastic, it is extremely important for system operators to employ algorithms by which all possible feasible equilibria of an overall complex power system model can be computed, and their stability properties can be evaluated so that power dispatches and control decisions can be planned accordingly.  The resulting operating
point as a solution of a highly nonlinear system of equations is known as the load flow problem \cite{kundur}. Computing equilibria by solving load flow equations for a conventional power system model has been a standard practice in the power industry for decades, with several advanced numerical algorithms being proposed in the literature \cite{hiskens, chiang, overbye}. However, there is still a lack of insight on how these equilibria characteristics may change as operators move towards more renewable penetration, especially the penetration of wind power which is scheduled to reach a $20\%$ penetration by 2030 \cite{wind}. Conventional load flow algorithms that generally estimate local solutions of nonlinear equations may not be sufficient to compute all possible feasible solutions in such cases, and more advanced algorithms are needed. 

In this paper, we address this problem and employ a homotopy based numerical continuation method to solve for power system equilibria with increasing levels of wind penetration. We first derive the end-to-end dynamic model of the wind-integrated system considering synchronous generators, loads and a wind plant consisting of a doubly-fed induction generator coupled to a wind turbine. The models of each of these subcomponents are coupled to each other via power flow, thereby leading to a differential-algebraic (DAE) model for the overall system. On applying the homotopy-based algorithm we find that even for fixed levels of power consumption and generation multiple solutions of the load flow equations can coexist, especially due to the strong influence of the voltage control setpoints in the DFIGs on the active and reactive power flows in the grid. This is, of course, in sharp contrast to the conventional load flow results where generally only one feasible or normal solution exists, characterized by high voltage level and correspondingly low currents \cite{konsta}. The method, therefore, brings out the explicit possibilities where increasing the level of wind penetration, and that too at specific buses in a system, may lead to very different, new stable solutions for load flow. 

In this paper, we introduce a {\it parameter homotopy} algorithm which can be viewed as an extension of the recently developed Numerical Polynomial Homotopy Continuation (NPHC) method \cite{sommese2005numerical}. The homotopy-based algorithm starts by observing that the power flow equations of the wind-integrated model are given by multivariate nonlinear polynomials by expanding the voltage phasor at any bus to its real and imaginary parts. Solving nonlinear algebraic equations, in general, is a highly nontrivial task. Using recently developed methods in computational algebraic geometry, however, deriving all solutions of a system of polynomial equations is now possible. One approach that is known to have only isolated solutions is to compute a Grobner basis (GB) that converts the original system of equations to a triangular form, similar to the row-echelon form for linear equations. The new system is referred to as a GB, and the algorithm to compute it as the Buchberger algorithm \cite{Montes:02,buchberger2006bruno,CLO:07,nguyen2014appearance}. Since complex solutions also include real solutions, by this approach one can find all  real solutions. So far the approach has been rather underutilized for load flow calculations in power systems except for a few attempts that are mostly based on small toy models \cite{70552,ma1993efficient,liu2005toward,molzahn2013counterexample,rostalski2011numerical}. The proposed method in this work can solve a set of nonlinear algebraic equations over a number of parameter values with greater computational efficiency as compared to the existing methods. Instead of solving the equations for each parameter value, the proposed method can solve the equations at generic complex values of the parameter. The solutions of the generic system can be used to obtain the solutions for the physical values of the parameter which drastically reduces the computational effort and time. Applying this method to a 7-bus 5-machine wind power system model we show that with increasing wind penetration the system may exhibit undesired or poorly damped poles, and in some instances even unstable poles.

The rest of the paper is organized as follows. In Section II we provide the dynamic model equations for synchronous generators, wind turbines and DFIG, and derive the coupled DAE model via power flow. In Section III we formulate the problem of equilibrium analysis. Section IV introduces the homotopy based approach to solve the nonlinear load flow model. Section V presents a case study on a representative power system model with considerable wind injection followed by the conclusions.

\section{A wind-integrated power system model}
 \label{sec:grid_model}
 We consider a power system with $N$ buses and $n$ generators. These generators consist of a set $\mathcal{G}=:\{1,\dots,n-1\}$ of synchronous generators and one wind power plant. Without loss of generality we can reorder the buses and classify them into 3 sets namely the set of synchronous generator bus $\mathcal{N}_{s}=:\{1, \dots,n-1\}$, the wind injection bus $\mathcal{N}_{w}=:\{n\}$ and the load buses $\mathcal{N}_{l}=:\{n+1, \dots,N\}$. To obtain the equilibrium for this system, our first task is to derive its dynamic model considering both nonlinear swing dynamics and wind power dynamics. This is presented as follows.
\subsection{Synchronous generator model} 
 We model each generator $i\in\mathcal{G}$ using the swing equations, 
\begin{subequations}
\label{eqn:sync_gen}
\begin{align}
{{{\dot \delta }_i}}&={{\omega _i}}\label{eqn:sync_gen_a}\\
m_{i} \dot{\omega _i} &= P_{mi}-P_{si}\label{eqn:sync_gen_b}.
\end{align}
\end{subequations}
Here $\delta_i$, $\omega_i$, $m_{i}$, $P_{mi}$ and $P_{si}$ are respectively the phase angle, rotor speed, inertia, the mechanical power input and the active power output of synchronous generator $i\in\mathcal{G}$. The active and reactive power output of the synchronous generator
$i\in\mathcal{G}$ can be respectively shown as,
\begin{subequations}
\label{eqn:syn_power}
\begin{align}
{P_{si}} =& \frac{E_{i}}{x_{di}^{'}}\left( {{V_{si_{Re}}}\sin {\delta_{i}} - {V_{si_{Im}}}\cos {\delta_{i}}}\right)\label{eqn:syncmc_active_i}\\
\label{eqn:syncmc_reactive_i}{Q_{si}} =& \frac{{E_i^2}}{{x_{di}^{'}}} - \frac{{{E_i}}}{{x_{di}^{'}}}\left( {{V_{si_{Re}}}\cos {\delta _i } -{V_{si_{Im}}}\sin {\delta _i }} \right),
\end{align}
\end{subequations}
 where, $E_{i}$ is the internal voltage and $x_{di}^{'}$ is the direct-axis salient reactance of the machine $i\in\mathcal{G}$. $V_{si}=V_{si_{Re}}+jV_{si_{Im}}$ is the voltage at bus $i\in\mathcal{N}_s$. Also $V_{si_{Re}}=\left| {{V_{si}}} \right|\cos {\theta _{si}}$ and  $V_{si_{Im}}=\left| {{V_{si}}} \right|\sin {\theta _{si}}$, $\theta_{si}$ being the angle of the voltage at bus $i\in\mathcal{N}_s$. 
 The states of the synchronous generator $i\in\mathcal{G}$ are coupled with the states of other generators and of the wind power plant via power flow, with the bus voltage $V_{si}$ being the coupling variable as shown in the next subsections.
\subsection{Wind power plant model} 
For convenience of analysis, we model the wind power plant by a single representative wind turbine coupled to a DFIG. This turbine serves as an aggregate model of a group of turbines injecting power to the grid at a point of common coupling, as shown in Figure \ref{fig:wind_gen}. This common bus is indexed as $\mathcal{N}_{w}$. We adjust the power output level of the resulting `wind injection' through a parameter $\gamma$ which represent the number of wind turbines connected to the bus $\mathcal{N}_{w}$.  Usually wind generators are of various types, each of which comes with its own set of  controls. In this study we consider Type III wind turbines. The turbine and the DFIG are connected by a shaft and a gear box as shown in Figure \ref{fig:wind_gen}. The details of the model are derived as follows.
\subsubsection{Wind turbine model}
\begin{figure}[t]
            \centering
             \includegraphics[width=0.46\textwidth]{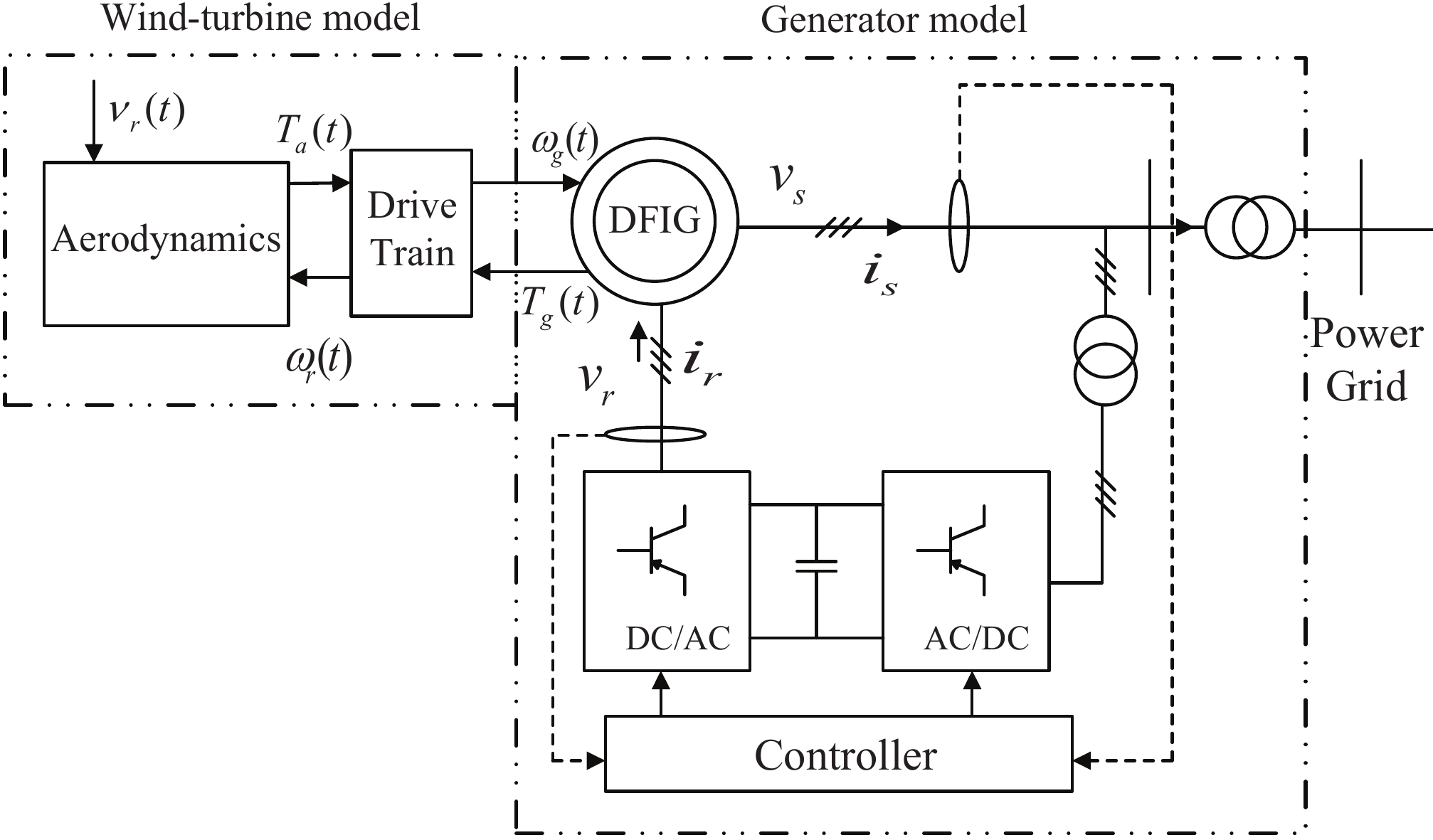}
             \caption{Wind turbine interfaced to the grid via a DFIG}
\label{fig:wind_gen}
\end{figure} 
The drive train for the wind turbine is modeled as two connected shafts operating at high and low speeds. The aerodynamic torque ${T_a}(t)$ of the wind turbine rotor is transferred via the shaft as the electromagnetic torque ${T_g}(t)$ driving the DFIG, as shown in Figure \ref{fig:wind_gen}. The expression for  ${T_a}(t)$ due to wind speed $v_{r}(t)$ is given as
\begin{equation}
\label{eqn:torq_aero}
T_a (t) = \frac{\rho {A_s}{(v_r (t))}^3{C_p (\lambda(t),\beta)}}{{2{\omega _r (t)}}}
\end{equation}
where, $A_s$ is the swept area of the turbine blades, $C_p$ is the power coefficient of the turbines, $\lambda (t)$ is the blade tip speed ratio, $\beta$ is the pitch angle, $\omega_{r} (t)$ is the speed of the turbine rotor, $\rho$ is the air density. The power coefficient $C_p$ is defined as,
\begin{equation}
\label{eqn:power_coeff}
{C_p}(\lambda (t),\beta) = 0.22\left( {\frac{{116}}{{{\lambda _i(t)}}} - 0.4\beta  - 5} \right){e^{ - \frac{{12.5}}{{{\lambda _i(t)}}}}}
\end{equation}
where,
${\lambda _i}(t) = {1}/{\left[1/\left( {\lambda(t)  + 0.08\beta} \right) - 0.035/\left( {{\beta ^3} + 1} \right)\right]}$, $\lambda(t)  = \omega _r (t) R/v_r (t)$, $R$ being the length of the wind turbine blade. We assume that the turbine is operating above the rated wind speed and the pitch angle $\beta=0$ at all times.
 Considering the frictional losses on the shaft to be negligible, the electromagnetic torque of the DFIG is given by,
\begin{equation}
\label{eqn:shaft_torq}
T_{g}(t)=T_a (t). 
\end{equation}
The mechanical speed of the DFIG, $\omega_{g}(t)$ is related to the turbine rotor speed $\omega_{r}(t)$ via a gear ratio $N_g$ as,
\begin{equation}
\label{eqn:speed_gear}
\omega_{g}(t) = {N_g}\omega_{r}(t)
\end{equation}
\subsubsection{DFIG model}
The DFIG is modeled through the dynamics of its stator and rotor variables, expressed in a rotating $d-q$ reference frame as \cite{ugalde2013},
\begin{subequations}\label{eqn:volt_cur}
\begin{align}
 \begin{split}
 v_{qs}(t)&= \left({{R_s} + \mathscr{D}{L_s}}\right)i_{qs}(t)+{{\omega _e}{L_s}}i_{ds}(t)\\
 &\quad\quad+{\mathscr{D}{L_m}}i_{qr}(t)+{{\omega _e}{L_m}}i_{dr}(t)\end{split}\\
 \begin{split}
  v_{ds}(t)&= { - {\omega _e}{L_s}}i_{qs}(t)+\left({{R_s} + \mathscr{D}{L_s}}\right)i_{ds}(t)\\
 &\quad\quad-{{\omega _e}{L_m}}i_{qr}(t)+{\mathscr{D}{L_m}}i_{dr}(t)\end{split}\\
 \begin{split}
v_{qr}(t)&= {\mathscr{D}{L_m}}i_{qs}(t)+{({\omega _e} - {\omega _{ge}}){L_m}}i_{ds}(t)\\
 &\quad+\left({{R_r} + \mathscr{D}{L_r}}\right)i_{qr}(t)+{({\omega _e} - {\omega
 _{ge}}){L_r}}i_{dr}(t)\end{split}\\
 \begin{split}
 v_{dr}(t)&= { - ({\omega _e} - {\omega _{ge}}){L_m}}i_{qs}(t)+{\mathscr{D}{L_m}}i_{ds}(t)\\
 &\quad{ - ({\omega _e} - {\omega _{ge}}){L_r}}i_{qr}(t)+\left({{R_r} + \mathscr{D}{L_r}}\right)i_{dr}(t),\end{split}
\end{align}
\end{subequations}
where $\omega_{ge}:=\frac{p}{2}\,\omega_{g}$ is the electrical speed, $p$ is the number of electrical poles of the DFIG, and $\mathscr{D}$ is the differential operator. The subscripts $d$ and $q$ refer to the direct and quadrature axes of the reference frame rotating at constant speed $\omega_e$. Subscripts $s$ and $r$ respectively indicate quantities associated with the stator and rotor circuits. The symbols $v$, $i$, and $R$ respectively denote voltage, current, and resistance. $L_{ls}$, $L_{lr}$, $L_m$ are respectively the stator and rotor leakage inductances, and the magnetizing inductance.   
The electromagnetic torque of the DFIG, shown in \eqref{eqn:shaft_torq} is given by
\begin{equation}
\label{eqn:Tem} {T_g}(t) = -\frac{{3p}}{4}{L_m}\left[
{{i_{qs}(t)}{i_{dr}(t)} - {i_{ds}(t)}{i_{qr}(t)}} \right].
\end{equation}
The active and reactive power output of the DFIG are,
\begin{subequations}
\label{eqn:im_power}
\begin{align}
 {P_{w}}  &=  {v_{qs}}{i_{qs}} +
{v_{ds}}{i_{ds}} + {v_{dr}}{i_{dr}} + {v_{qr}}{i_{qr}}
\label{eqn:im_active}\\
 {Q_{w}} &= {v_{ds}}{i_{qs}} -
{v_{qs}}{i_{ds}}+{v_{dr}}{i_{qr}} - {v_{qr}}{i_{dr}}.\label{eqn:im_reactive}
\end{align}
\end{subequations}
The stator of the DFIG is directly connected to the wind injection bus. The $q$ and $d$ directions in our analysis are so chosen that $v_{qs}$ and $v_{ds}$ align with $V_{wi_{Re}}$ and $V_{wi_{Im}}$, the real and imaginary parts of the voltage $V_{wi}$ of bus $i\in\mathcal{N}_{w}$, respectively. $V_{wi}$, therefore, serves as the coupling variable for the DFIG states to the rest of the system. We next derive the power flow equations for this wind integrated system.

\subsection{Power flow model} 
The dynamics of the synchronous machines, the wind power plant and the loads in the power system are connected by the power flow equations between the different buses. The active and reactive power flows for each bus are respectively given by
\begin{subequations}
\label{eqn:load_flows}
\begin{align}
\label{eqn:load_active} 0=&{P_{j}} - {\rm{Re}}\left\{
{\sum\limits_{k = 1,k \ne j}^N { {{V_{jk}}}{{\left(
{\frac{{{V_j}}}{{{Z_{jk}}}}} \right)}^*}} } \right\} -| V_j|^2{G_j}\\
\label{eqn:load_reactive} 0=&{Q_{j}} - {\mathop{\rm
Im}\nolimits} \left\{ {\sum\limits_{k = 1,k \ne j}^N
V_{jk}{{{\left( {\frac{{{V_{j}}}}{{{Z_{jk}} }}} \right)}^*}} }
\right\} -| V_j|^2{B_j}.
\end{align}
\end{subequations}
Here $P_{j}$ and $Q_{j}$ respectively denote the active and reactive power flow from the generators connected to bus $j$, $G_j$ and $B_j$ are respectively the load conductance and load suceptance with line charging of bus $j$. $Z_{jk}$ is the impedance of the lines connecting buses $j$ and $k$, and $V_{jk}=\left(V_{j_{Re}}-V_{j_{Re}}\right)+j\left(V_{k_{Re}}-V_{k_{Im}}\right).$ For any bus $j\in\mathcal{N}_{s}$,  $P_{j}$ and $Q_{j}$ are respectively equal to $P_{sj}$ and $Q_{sj}$ shown in \eqref{eqn:syn_power}. For any bus $j\in\mathcal{N}_{w}$,  $P_{j}$ and $Q_{j}$ are respectively equal to $P_{wj}$ and $Q_{wj}$ shown in \eqref{eqn:im_power}. If $j\in\mathcal{N}_{l}$,  $P_{j}$ and $Q_{j}$ are both equal to $0$. The overall model of the wind integrated power system is, thus, comprised of the differential-algebraic equations  \eqref{eqn:sync_gen}-\eqref{eqn:load_flows}. Next we analyze the equilibria for this model. 
\section{Equilibrium Analysis}
 \label{sec:equ_anal}
In this section we derive the equilibrium or the steady-state values of all the dynamic variables in the wind integrated power system \eqref{eqn:sync_gen}-\eqref{eqn:load_flows}.. The superscript $e$ for any variable from now onwards will be used to indicates its equilibrium value(s). First, a steady-state power flow problem is solved for each bus in the power system. One of the synchronous generator bus $i\in\mathcal{N}_{s}$ is assumed to be the slack bus, for which $\left| {{V_{si}^e}} \right|$ and  ${{\theta_{si}^e}}$ are respectively equal to 1 and 0. The active and reactive power flow, ${P_{si}^e}$ and ${Q_{si}^e}$ at the slack bus at steady-state can be obtained by solving,
 \begin{subequations}
 \label{eqn:power_slbus}
\begin{align}
 {P_{si}^e} =&{\rm{Re}}\left\{
{\sum\limits_{k = 1,k \ne i}^N { {\left(1-V_{k}^e \right)}{{\left(
{\frac{{1}}{{{Z_{ik}}}}} \right)}^*}} } \right\} + {G_i}\\
 {Q_{si}^e}=& {\mathop{\rm
Im}\nolimits} \left\{ {\sum\limits_{k = 1,k \ne i}^N
\left(1-V_{k}^e \right){{{\left( {\frac{{{1}}}{{{Z_{ik}} }}} \right)}^*}} }
\right\} + {B_i}.
\end{align}
\end{subequations}
For all other synchronous generator bus $j\in\mathcal{N}_{s}$ such that $j\neq i$,  $\left| {{V_{sj}^e}} \right|$ and  ${P_{sj}^e}$ are known from the power system specifications. The variables ${{\theta_{sj}^e}}$ and  ${Q_{sj}^e}$ are obtained by solving
\begin{subequations}
\label{eqn:power_synbus}
\begin{align}
 {P_{sj}^e}=& {\rm{Re}}\left\{
{\sum\limits_{k = 1,k \ne j}^N { {{V_{jk}^e}}{{\left(
{\frac{{{V_{sj} ^e}}}{{{Z_{jk}}}}} \right)}^*}} } \right\} + {|V_{sj}^e|}^2{G_j}\\
 {Q_{sj}^e} =& {\mathop{\rm
Im}\nolimits} \left\{ {\sum\limits_{k = 1,k \ne j}^N
V_{jk}^e{{{\left( {\frac{{{V_{sj}^e}}}{{{Z_{jk}} }}} \right)}^*}} }
\right\} + {|V_{sj}^e|}^2{B_j}.
\end{align}
\end{subequations}
For the wind bus $j\in\mathcal{N}_{w}$, the connected wind generators are of Type III in which the steady state power output $P_{wj}^e$ and the turbine speed $\omega_{r} ^{e}$ are specified via maximum power point tracking algorithm from the turbine characteristics curve, for a given wind speed $v_{r}^e$ and coefficient $C_p ^e$.  We assume that the magnitude of the voltage at the wind injection bus $N_w$, denoted as $\left|V_{wj}^e\right]$ is controlled internally to a given value. The variables ${{\theta_{wj}^e}}$ and  ${Q_{wj}^e}$ are obtained by solving
\begin{subequations}
\label{eqn:power_windbus}
\begin{align}
{P_{wj}^e}=& {\rm{Re}}\left\{
{\sum\limits_{k = 1,k \ne j}^N { {{V_{jk}^e}}{{\left(
{\frac{{{V_{wj} ^e}}}{{{Z_{jk}}}}} \right)}^*}} } \right\} + {|V_{wj}^e|}^2{G_j}\\
 {Q_{wj}^e}=&{\mathop{\rm
Im}\nolimits} \left\{ {\sum\limits_{k = 1,k \ne j}^N
V_{jk}^e{{{\left( {\frac{{{V_{wj}^e}}}{{{Z_{jk}} }}} \right)}^*}} }
\right\}+{|V_{wj}^e|}^2{B_j}.
\end{align}
\end{subequations}
However, the stator voltages of the DFIG $v_{qs}^e$ and $v_{ds}^e$ are respectively equal to $V_{{wj}_{Re}}$ and $V_{{wj}_{Im}}$. So the steady-state equations of the wind generators also need to be solved simultaneously with those for the synchronous generators. These are given as
\small
\begin{subequations}\label{eqn:power_windgen}
\begin{align}
{T_g}^e &= -\left({3p}/{4}\right){L_m}\left[{{i_{qs}^e}{i_{dr}^e} - {i_{ds}^e}{i_{qr}^e}} \right]\\
 v_{qs}^e&= {{R_s} }{i_{qs}^e}+{{\omega _e}{L_s}}{i_{ds}^e}+{{\omega _e}{L_m}}{i_{dr}^e}\\
 v_{ds}^e&= { - {\omega _e}{L_s}}{i_{qs}^e}+{{R_s}}{i_{ds}^e}-{{\omega _e}{L_m}}{i_{qr}^e}\\
 \begin{split}
 v_{qr}^e&= {({\omega _e} - {\omega _{ge}}){L_m}}i_{ds}^e +{{R_r}}i_{qr}^e+{({\omega _e} - {\omega_{ge}}){L_r}}i_{dr}^e\end{split}\\
 \begin{split}
 v_{dr}^e&= { - ({\omega _e} - {\omega _{ge}}){L_m}}{i_{qs}^e}+{{R_r}}{i_{dr}^e}{ - ({\omega _e} - {\omega _{ge}}){L_r}}{i_{qr}^e}\end{split}\\
 {P_{wj}^e}  &= \gamma \left({v_{qs}^e}{i_{qs}^e} +
{v_{ds}^e}{i_{ds}^e} + {v_{dr}^e}{i_{dr}} + {v_{qr}^e}{i_{qr}^e}\right)\\
 {Q_{wj}^e} &= \gamma \left({v_{ds}^e}{i_{qs}^e} -
{v_{qs}^e}{i_{ds}^e}+{v_{dr}^e}{i_{qr}^e} - {v_{qr}^e}{i_{dr}^e}\right).
\end{align}
\end{subequations}
\normalsize
For each load bus $j\in\mathcal{N}_{l}$, the bus voltage magnitude $\left|V_{{lj}^e}\right]$ and angle ${{\theta_{lj}^e}}$ are obtained by solving
\begin{subequations}
\label{eqn:power_loadbus}
\begin{align}
 0=& {\rm{Re}}\left\{
{\sum\limits_{k = 1,k \ne j}^N { {{V_{jk}^e}}{{\left(
{\frac{{{V_{lj} ^e}}}{{{Z_{jk}}}}} \right)}^*}} } \right\} +{|V_{lj}^e|}^2{G_j}\\
 0=& {\mathop{\rm
Im}\nolimits} \left\{ {\sum\limits_{k = 1,k \ne j}^N
V_{jk}^e{{{\left( {\frac{{{V_{lj}^e}}}{{{Z_{jk}} }}} \right)}^*}} }
\right\} +{|V_{lj}^e|}^2{B_j}.
\end{align}
\end{subequations}
The complete equilibria of the wind-integrated power system can be obtained by solving the set of nonlinear algebraic equations shown in \eqref{eqn:power_slbus}-\eqref{eqn:power_loadbus}. These equations depend upon the the amount of the wind penetration specified by $\gamma$, and the set point  $\left|V_{wj}^e\right|$ for controlling the voltage level of the wind bus. In other words, the equilibrium points of the system are parameterized by $\gamma$ and $\left|V_{wj}^e\right|$. In the next section we introduce a homotopy-based numerical continuation method by which we can solve for all such feasible equilibria.

\section{Solving DAE Equilibrium via Homotopy Continuation Method}
\label{sec:numer_homotop}
In this section we solve the equilibrium of the wind integrated power system shown in \eqref{eqn:power_slbus}-\eqref{eqn:power_loadbus} using the homotopy based continuation method. First we express these equations in polynomial form by replacing the complex bus voltages $V_{sj}^{e}$,  $V_{wj}^{e}$ and $V_{lj}^{e}$ in terms of their real and imaginary components. Also the corresponsding angles of the bus voltages $\theta_{sj}^{e}$, $\theta_{wj}^{e}$ and $\theta_{lj}^{e}$ are expressed in terms of their sine and cosine components. Additional equations are required to pose the problem in a polynomial form as  additional variables are introduced in the process.  Equations of the form, $\sin^{2}\theta+\cos^{2}\theta=1$, where $\theta$ is any bus voltage angle, are used together with the power system equations for this purpose.
In \eqref{eqn:power_slbus}-\eqref{eqn:power_loadbus}, there exist two main parameters, namely the level of wind penetration denoted as $\gamma$ and the setpoint for the voltage controller in the DFIG, denoted as $\left|V_{wj}^e\right|$, with respect to which their solutions can be tested. One way to achieve that will be to solve the equations for every feasible combinations of the parameter values, which is highly inefficient
and computationally expensive. One can in principle resort to the so-called comprehensive GB technique in which one can leave the parameters in the
symbolic form. 
This idea was used in \cite{Montes:02} with an attempt to use the CGB technique to solve parametric power flow for conventional grid models. More recently, a numerical method called the numerical polynomial homotopy continuation (NPHC) method \cite{sommese2005numerical} 
has successfully been applied to solve up to $14$ bus power flow systems in \cite{konsta}. The method, like the GB method, is based on complex algebraic geometry and 
guarantees to find \textit{all} the complex, and hence, real solutions of the system. We recall that for a system of polynomial equations without parameters
\begin{figure}[t]
            \centering
             \includegraphics[width=0.4\textwidth]{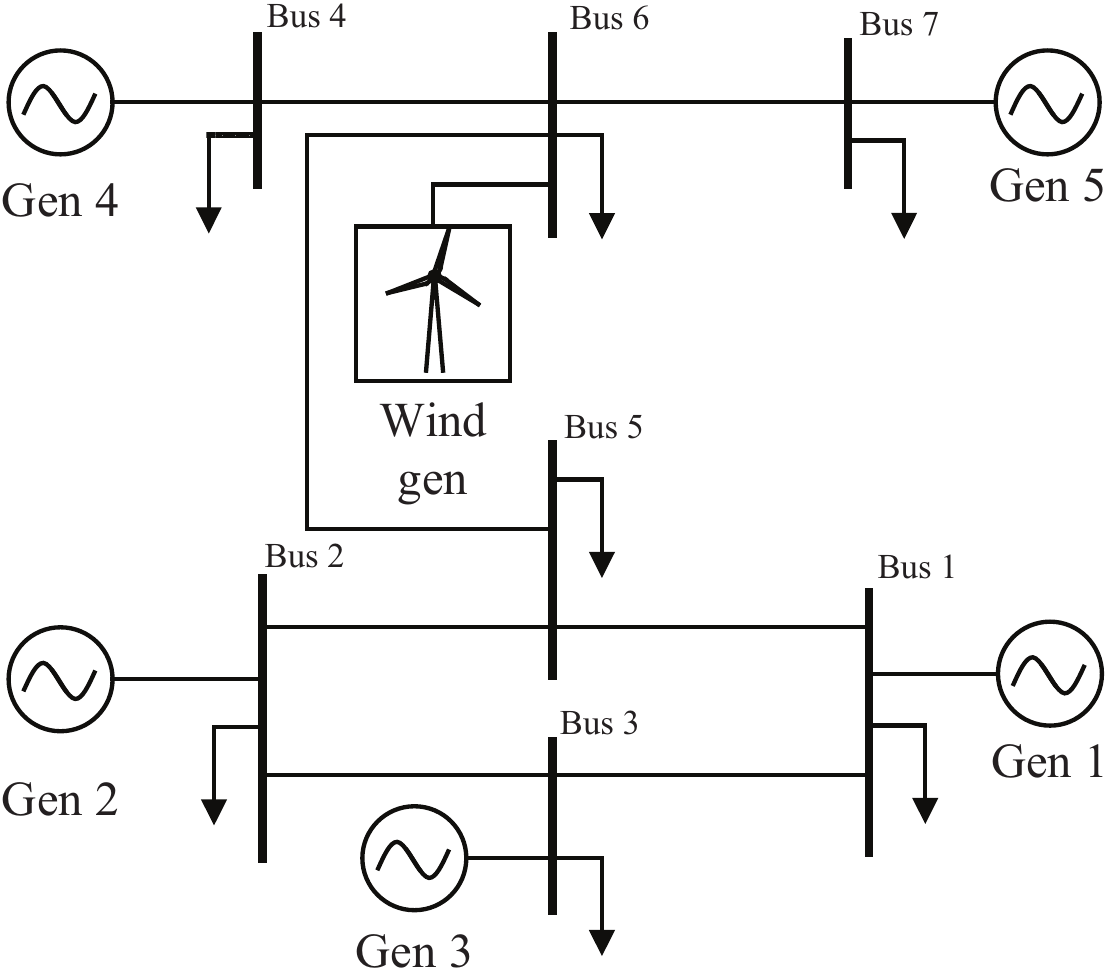}
             \caption{The 7-bus, 5 machine Brazilian system with wind injection}
\label{fig:grid_7bus}
\end{figure} 
\begin{equation}
P(x)=(p_{1}(x),\dots,p_{m}(x))^{T}=0, \;\; x=(x_{1},\dots,x_{m})^{T}
\end{equation}
\begin{table*}[t]
\centering
\begin{tabular}{| p{0.1\textwidth} |p{0.15\textwidth}|p{0.15\textwidth}|p{0.15\textwidth}| p{0.15\textwidth}|}
\hline
•\centering & \centering$\gamma=0.5$&\centering$\gamma=1.0$&\centering$\gamma=1.5$&\qquad\quad$\gamma=2.0$\\ 
\hline
\centering  $\left|V_{w6}^e\right|=0.96$ & \centering $\text{2 stable equilibria}$ &
 \centering $\text{2 stable equilibria}$ &  \centering $\begin{aligned} &\text{1 stable equilibrium,}\\ &\text{1 unstable equilibrium}\end{aligned}$ &  $\text{2 unstable equilibria}$ \\ 
\hline 
\centering  $\left|V_{w6}^e\right|=0.98$ & \centering $\text{2 stable equilibria}$ &
 \centering $\text{2 stable equilibria}$ &  \centering $\text{2 stable equilibria}$ &  $\text{2 unstable equilibria}$ \\  
\hline
\centering  $\left|V_{w6}^e\right|=1.00$ & \centering $\text{2 stable equilibria}$ &
 \centering $\text{2 stable equilibria}$ &  \centering $\text{2 stable equilibria}$ &  $\begin{aligned} &\text{1 stable equilibrium,}\\ &\text{1 unstable equilibrium}\end{aligned}$ \\ 
\hline
\end{tabular} 
\caption{Types of feasible equilibrium for varying wind penetration ($\gamma$) and wind bus voltage level  $\left|V_{w6}^e\right|$}\label{table:stability}
\end{table*}
with only isolated solutions, the classical B\'ezout Theorem asserts that for generic values of coefficients, 
the maximum number of isolated solutions in $\mathbb{C}^{m}$ is $\prod_{i=1}^{m}d_{i}$, where $d_{i}$ is the degree of the $i^{th}$ polynomial. Based on this bound, a homotopy $H(x,t)$ can be constructed as
\begin{equation}
H(x,t)=\gamma_{h}(1-t)Q(x)+t\; P(x),
\end{equation}
where $Q(x)=(q_{1}(x),\dots,q_{m}(x))^{T}$  is another system of polynomial equations, called the start system, which is created so that: 
\begin{enumerate}
\item the solutions of $Q(x)=H(x,0)=0$ are known or 
can be easily obtained
\item the number of solutions of $Q(x)=H(x,0)=0$ is equal to the CBB;
\item the solution set of $H(x,t)=0$ for $0\le t\le 1$ consists of a finite
number of smooth paths parameterized by
$t\in[0,1)$;
\item every isolated solution of $H(x,1)=P(x)=0$ can be obtained by some
path originating at a solution of $H(x,0)=Q(x)=0$.
\end{enumerate}
Hence, one then needs to track each solution of $H(x,0)=Q(x)=0$ from $t=0$ to $t=1$ using an efficient predictor-corrector method. 
Some paths may diverge to infinity along the way, but 
some paths which reach $P(x)=0=H(x,1)$ are \textit{all} the complex solutions of $P(x)=0$ as long as $\gamma_h$ is chosen to be a generic complex
number initially.

For a parametric system of polynomial equations, $P(x, \lambda) = 0$ where now $\lambda = (\lambda_1, \dots, \lambda_m) \in \mathbb{C}^m$
are parameters of the system, it can be shown that the maximum number of solutions 
over all the parameter-points is the one at a generic parameter-point \cite{li1989cheater,Morgan1989123}. 
 Algorithmically even though the parameters $\lambda$ in the systems coming from the power flow equations may be real, 
in the first step we need to solve $P(x, \lambda)$
at a \textit{generic} complex parameter-point $\lambda^{*} \in \mathbb{C}^{m}$, using the 
NPHC method with the CB bound. Then, in the second step, the system $P(x, \lambda^{*})=0$ becomes the start system for all other parameter-points 
$\lambda \in \mathbb{C}^{m} - \{ \lambda^{*} \}$ including the real ones, i.e., each solution of this start system 
needs to be tracked with the following homotopy:
\begin{equation}
H(x,\lambda, t)=(1-t)P(x, \lambda^{*}) + t\; P(x, \lambda).
\end{equation}
Here, the $\gamma_{h}$ trick is implicit for the generic $\lambda^*$. Usually,  the number of start solutions is significantly smaller 
than the CBB of the system reducing the number of paths to be tracked.  More importantly, one has to obtain 
the start system for a generic parametric system once for all, which can be done `offline'.   Then using this start system and solutions, one can obtain all the complex and real isolated solutions of the power flow model (\ref{eqn:load_flows}) at  as many values of $\gamma$ and $\left|V_{wj}^e\right|$in the parameter-space as required with a much faster `online' run. The method, therefore, converges very fast due to its inherent parallelization.
We next apply the above method to solve for all equilibria of \eqref{eqn:power_slbus}-\eqref{eqn:power_loadbus} for different combinations of $\gamma$ and $\left|V_{wj}^e\right|$, and test their stability properties.
 \section{Simulation Results}
 \label{sec:results}
 In this section we obtain all the equilibria of the wind-integrated model by solving \eqref{eqn:power_slbus}-\eqref{eqn:power_loadbus} by the homotopy based algorithm mentioned in Section \ref{sec:numer_homotop}. For the simulation we use the 7-bus, 5-machine equivalent model of the Southeastern Brazilian power system \cite{brazil7bus}, as shown in Figure \ref{fig:grid_7bus}. 
The parameters for the synchronous generators, the loads, the various line impedances of the power system, used in the simulation are taken from \cite{brazil7bus}. Bus 7 is assumed to be the slack bus with $\left| {{V_{s7}^e}} \right|$ and ${{\theta_{s7}^e}}$ respectively equal to 1 and 0. The bus voltage magnitudes of the other synchronous generator buses, i.e. $\left| {{V_{sj}^e}} \right|$ for $j\in\{1,2,3,4\}$ in ascending order of $j$ are given as $1.066,1.066,1.065 \text{ and } 1.076$. 

\begin{figure*}[t]
        \begin{subfigure}[b]{0.3\textwidth}
                \centering
                \includegraphics[width=\textwidth]{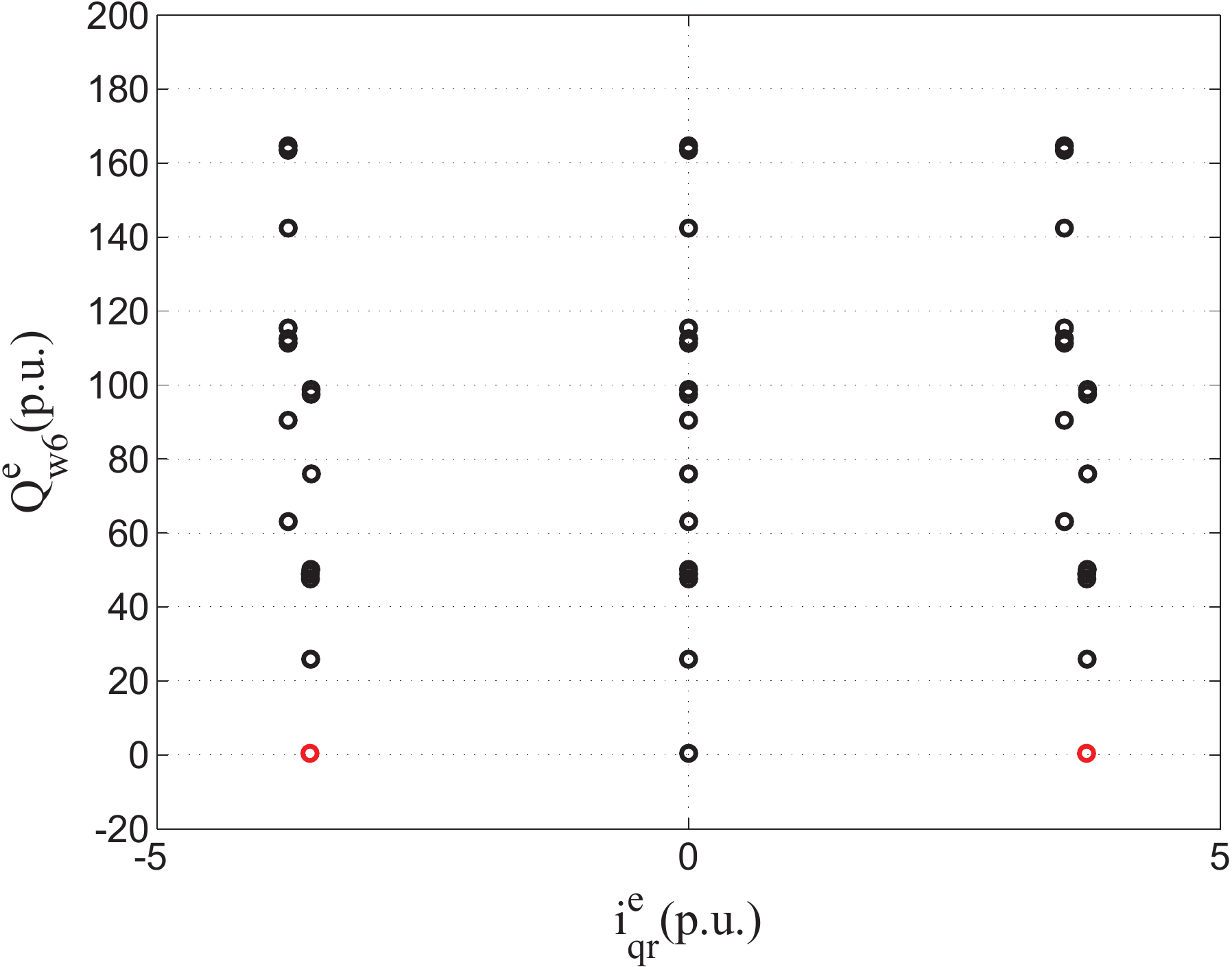}
                \caption{$\gamma$=0.5, $\left|V_{wj}^e\right]$=0.96 p.u.}
                \label{fig:grid_far}
        \end{subfigure}
          \quad
             \begin{subfigure}[b]{0.3\textwidth}
                \centering
                \includegraphics[width=\textwidth]{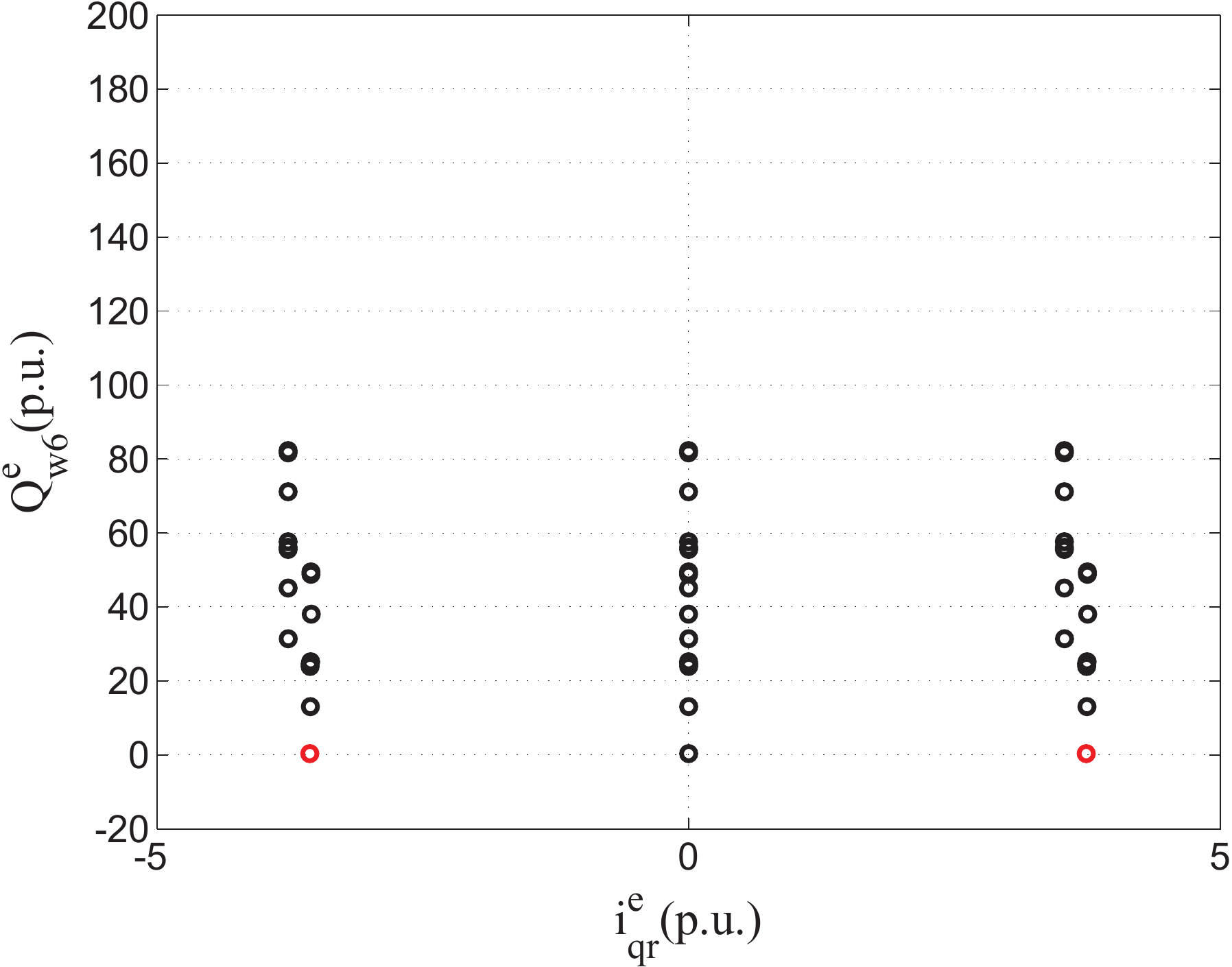}
                \caption{$\gamma$=1, $\left|V_{wj}^e\right]$=0.96 p.u.}
                \label{fig:angle_far}
        \end{subfigure}
        \quad
        \begin{subfigure}[b]{0.3\textwidth}
                \centering
                \includegraphics[width=\textwidth]{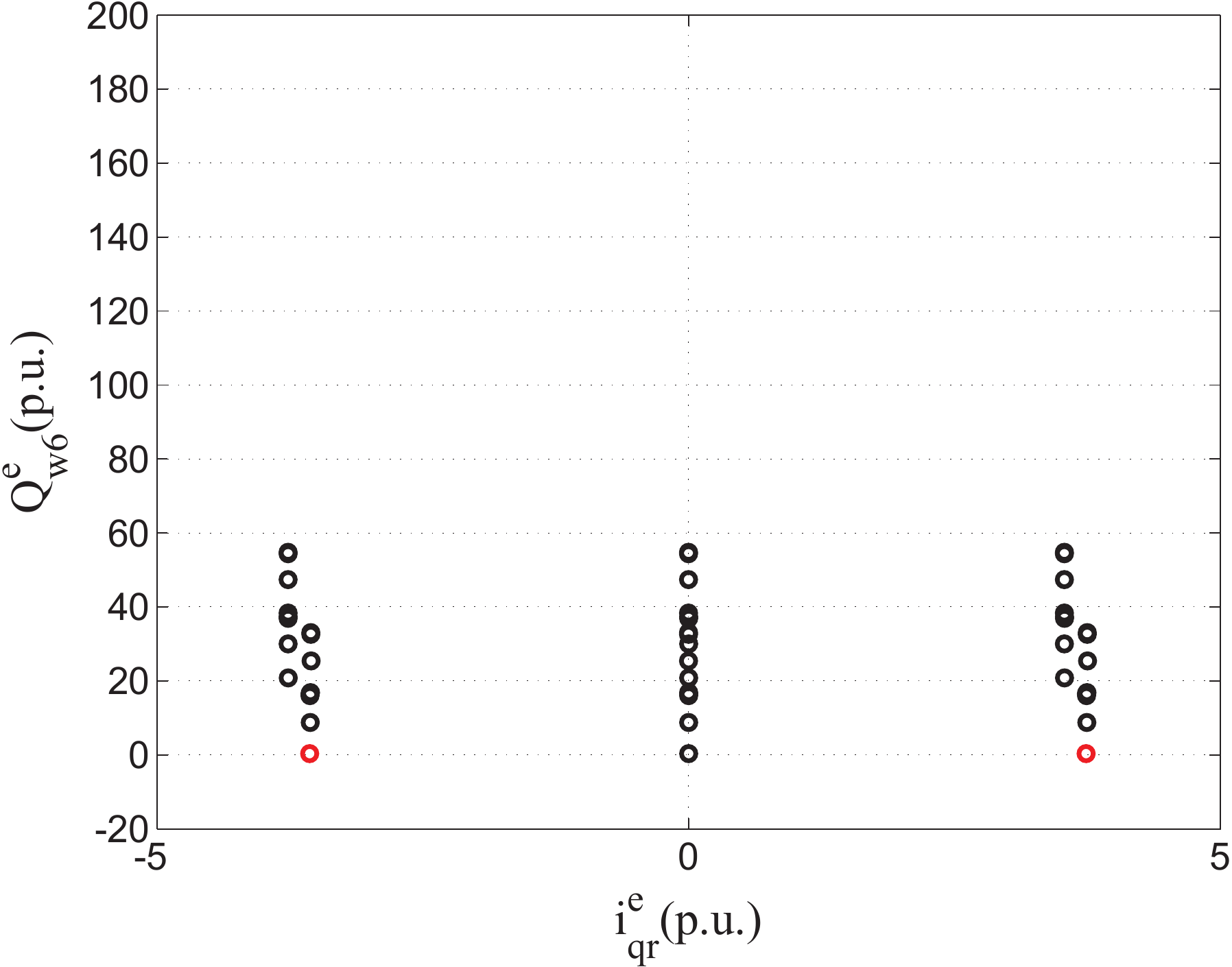}
                \caption{$\gamma$=1.5, $\left|V_{wj}^e\right]$=0.96 p.u.}
                \label{fig:mode_far}
        \end{subfigure}
\newline
\quad
 \begin{subfigure}[b]{0.3\textwidth}
                \centering
                \includegraphics[width=\textwidth]{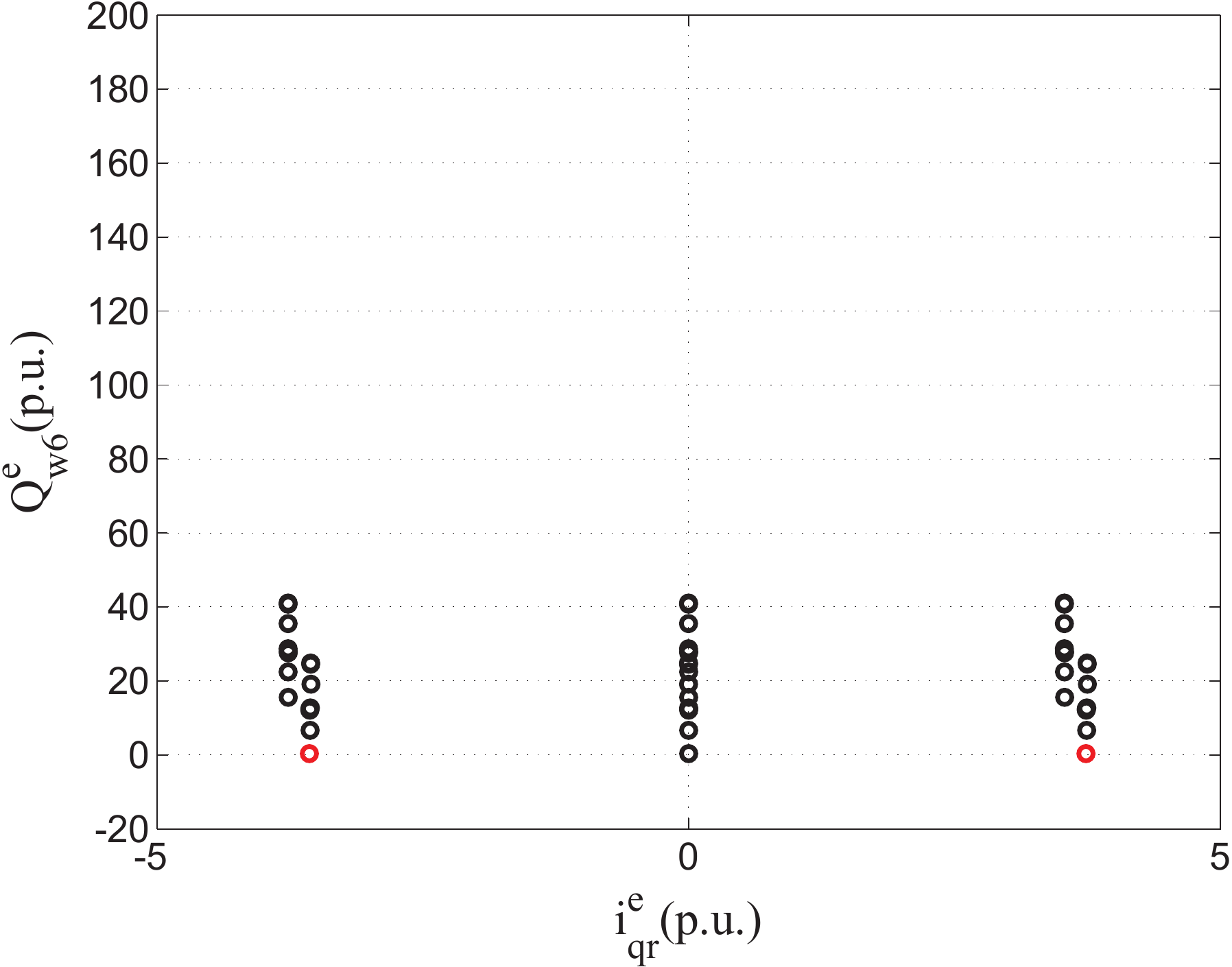}
                \caption{$\gamma$=2, $\left|V_{wj}^e\right]$=0.96 p.u.}
                \label{fig:grid_near}
        \end{subfigure}
          \quad
             \begin{subfigure}[b]{0.3\textwidth}
                \centering
                \includegraphics[width=\textwidth]{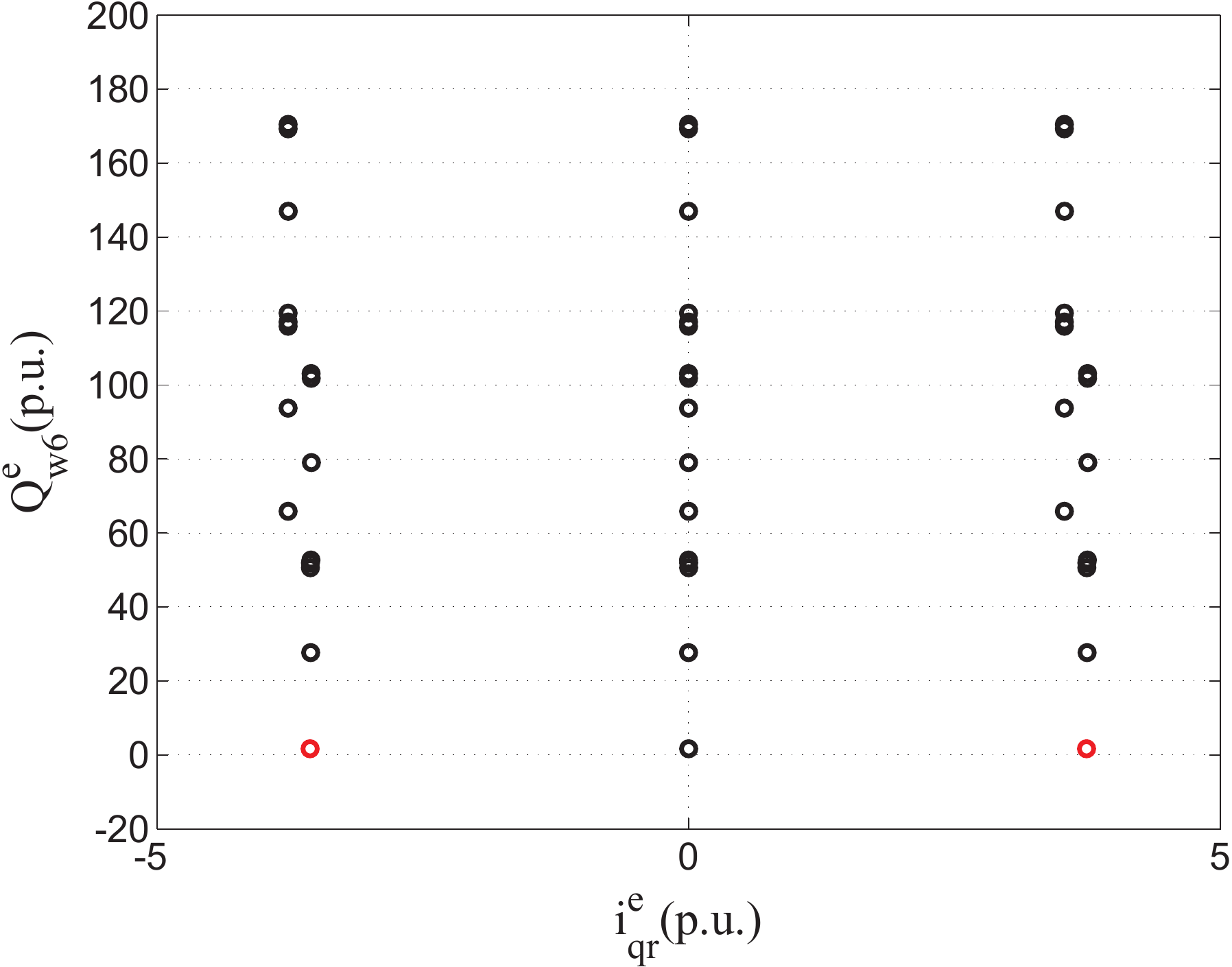}
                \caption{$\gamma$=0.5, $\left|V_{wj}^e\right]$=0.98 p.u.}
                \label{fig:angle_near}
        \end{subfigure}
           \quad
            \begin{subfigure}[b]{0.3\textwidth}
                \centering
                \includegraphics[width=\textwidth]{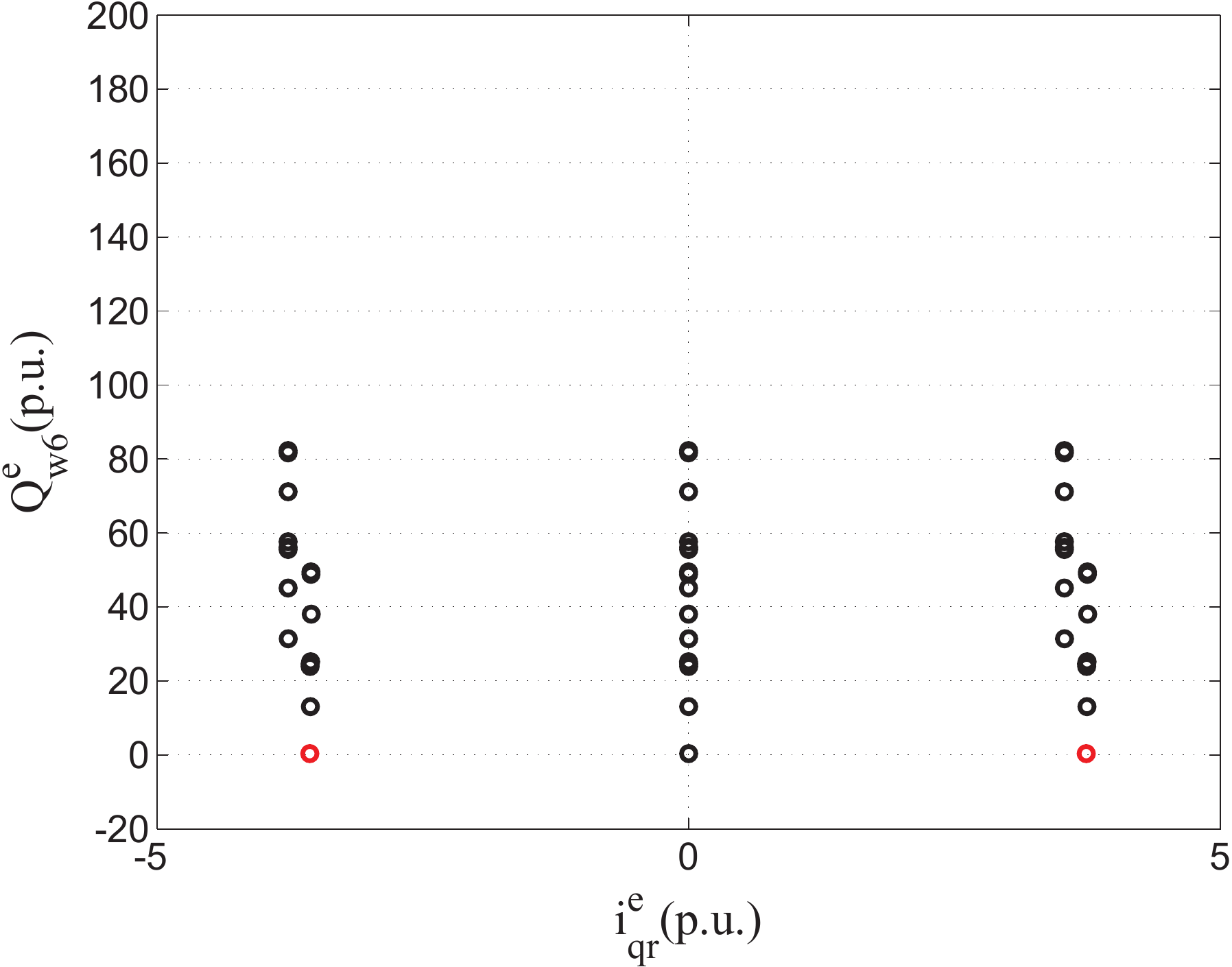}
                \caption{$\gamma$=1, $\left|V_{wj}^e\right]$=0.98 p.u.}
                \label{fig:mode_near}
        \end{subfigure}
\newline
\quad
 \begin{subfigure}[b]{0.3\textwidth}
                \centering
                \includegraphics[width=\textwidth]{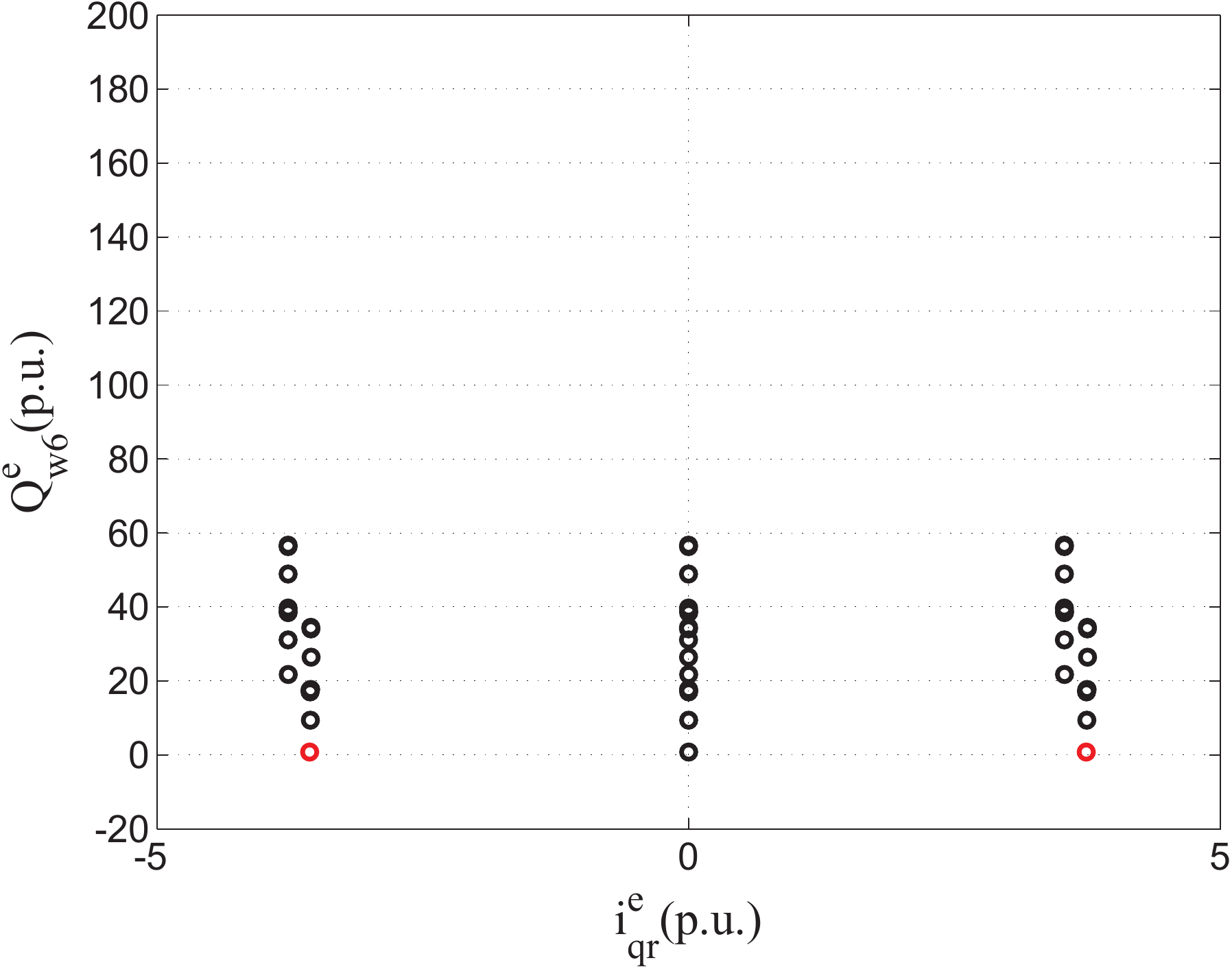}
                \caption{$\gamma$=1.5, $\left|V_{wj}^e\right]$=0.98 p.u.}
                \label{fig:grid_longwind}
        \end{subfigure}
          \quad
                \begin{subfigure}[b]{0.3\textwidth}
                \centering
                \includegraphics[width=\textwidth]{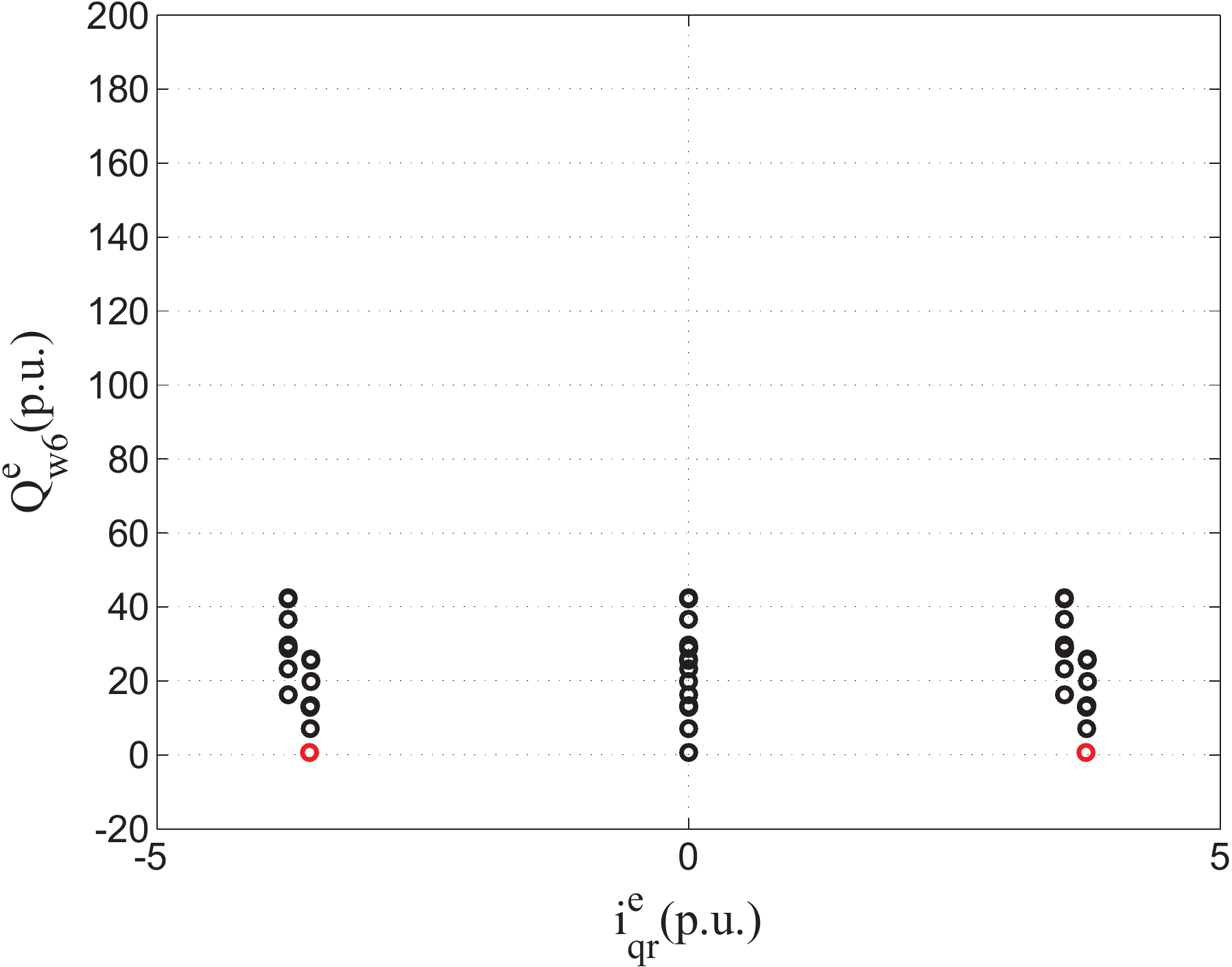}
                \caption{$\gamma$=2, $\left|V_{wj}^e\right]$=0.98 p.u.}
                \label{fig:angle_longwind}
        \end{subfigure}
        \quad
             \begin{subfigure}[b]{0.3\textwidth}
                \centering
                \includegraphics[width=\textwidth]{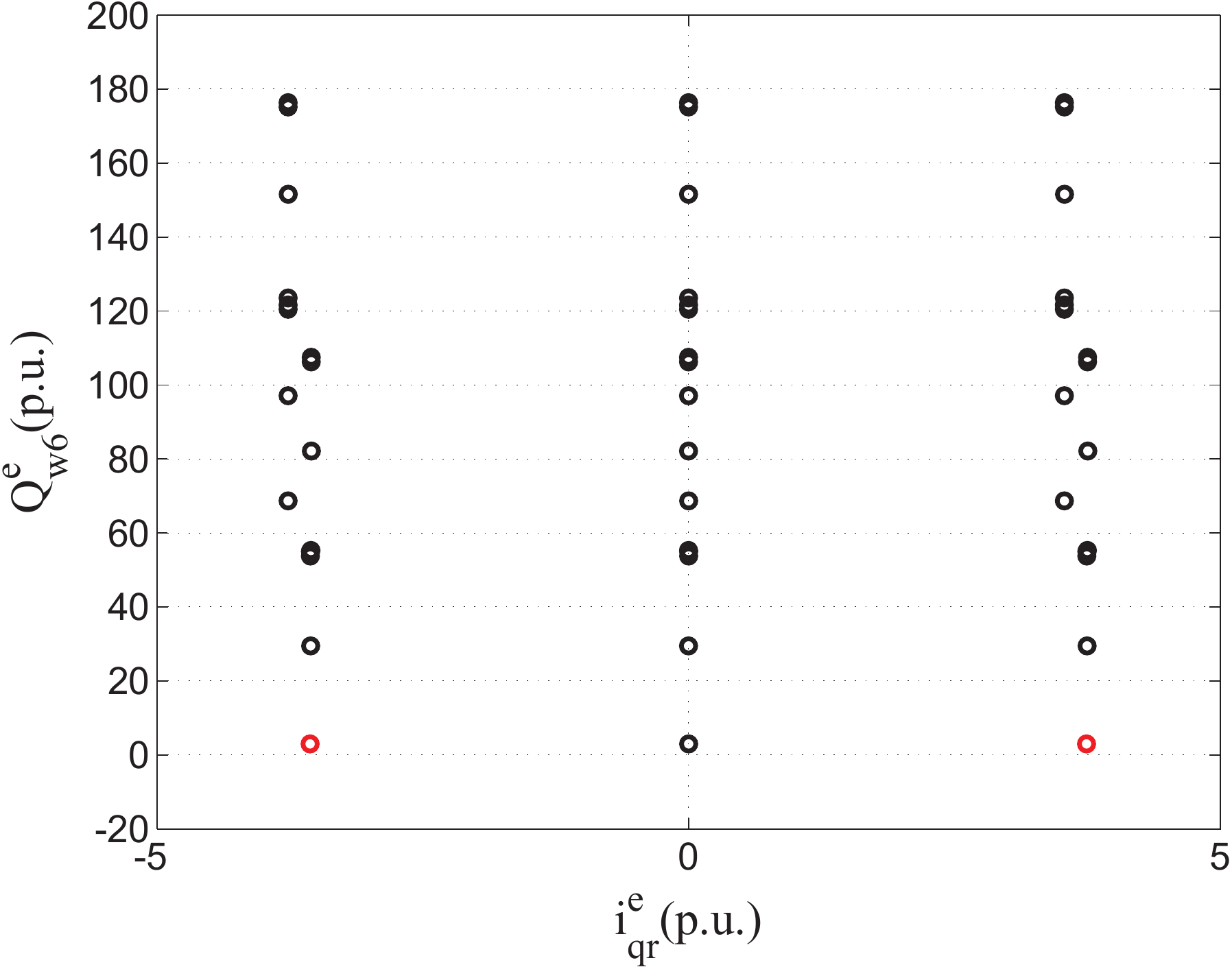}
                \caption{$\gamma$=0.5, $\left|V_{wj}^e\right]$=1 p.u.}
                \label{fig:mode_longwind}
        \end{subfigure}
\newline
\quad
 \begin{subfigure}[b]{0.3\textwidth}
                \centering
                \includegraphics[width=\textwidth]{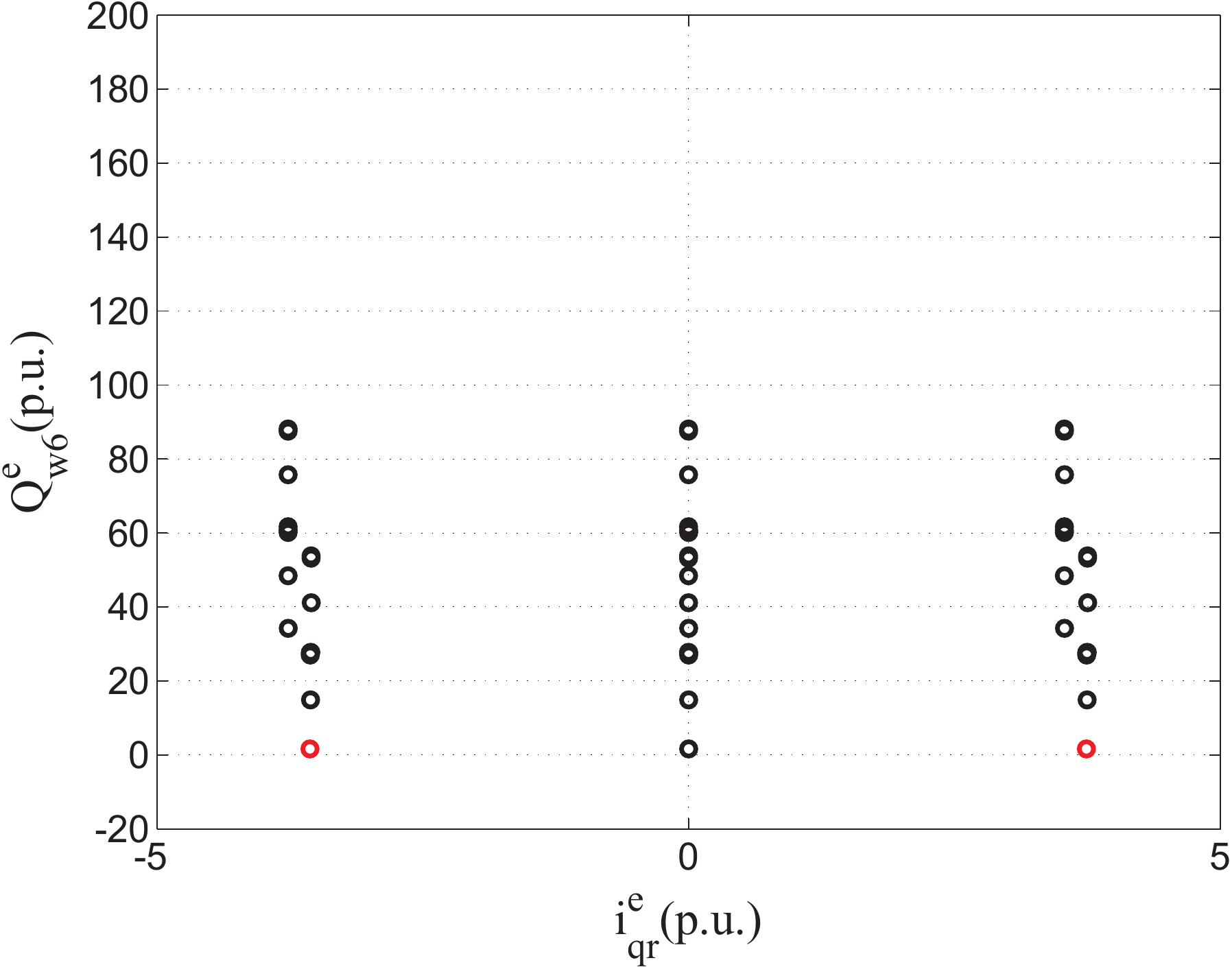}
                \caption{$\gamma$=1, $\left|V_{wj}^e\right]$=1 p.u.}
                \label{fig:grid_longwind12}
        \end{subfigure}%
          \quad
                \begin{subfigure}[b]{0.3\textwidth}
                \centering
                \includegraphics[width=\textwidth]{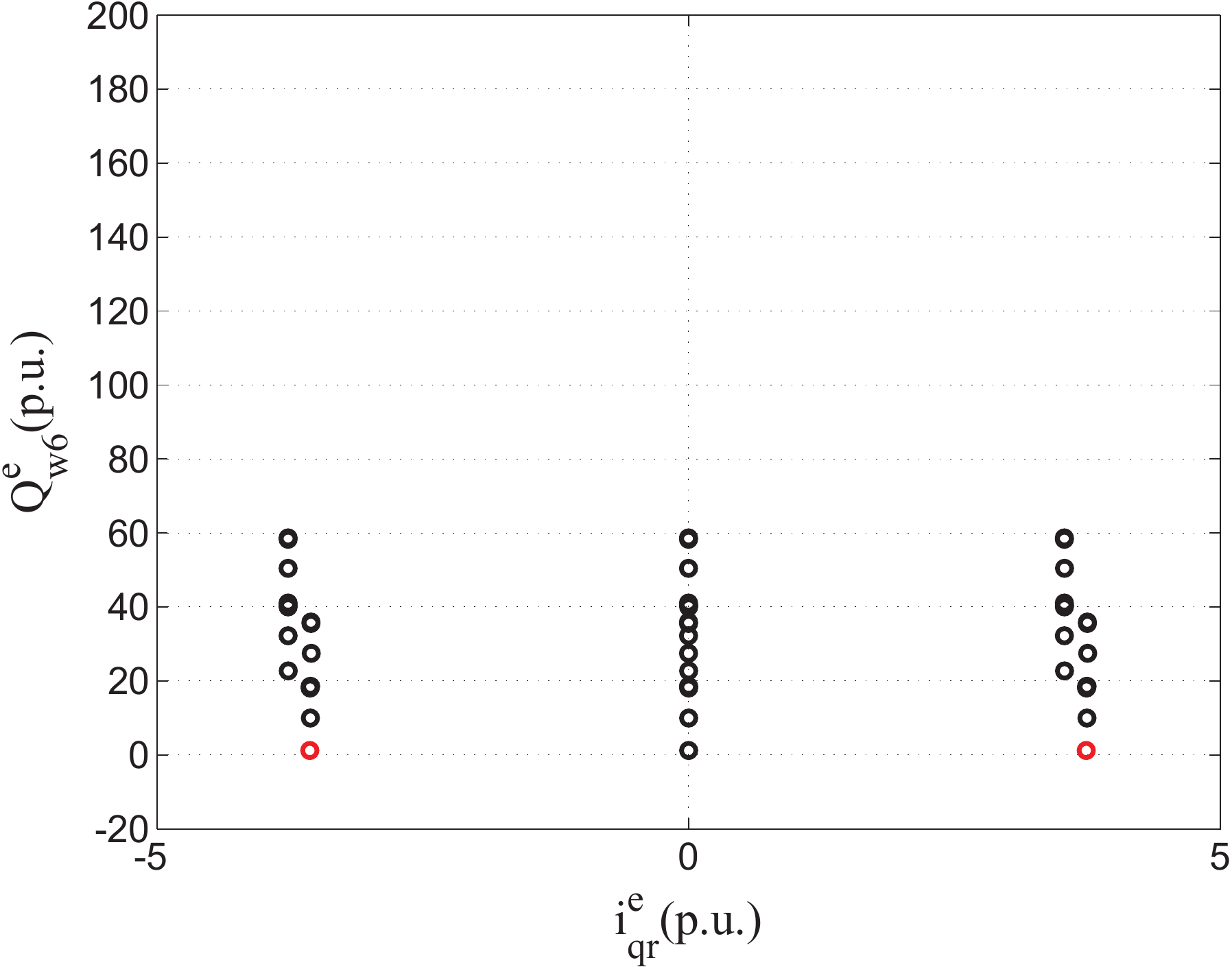}
                \caption{$\gamma$=1.5, $\left|V_{wj}^e\right]$=1 p.u.}
                \label{fig:angle_longwind12}
        \end{subfigure}
        \quad
             \begin{subfigure}[b]{0.3\textwidth}
                \centering
                \includegraphics[width=\textwidth]{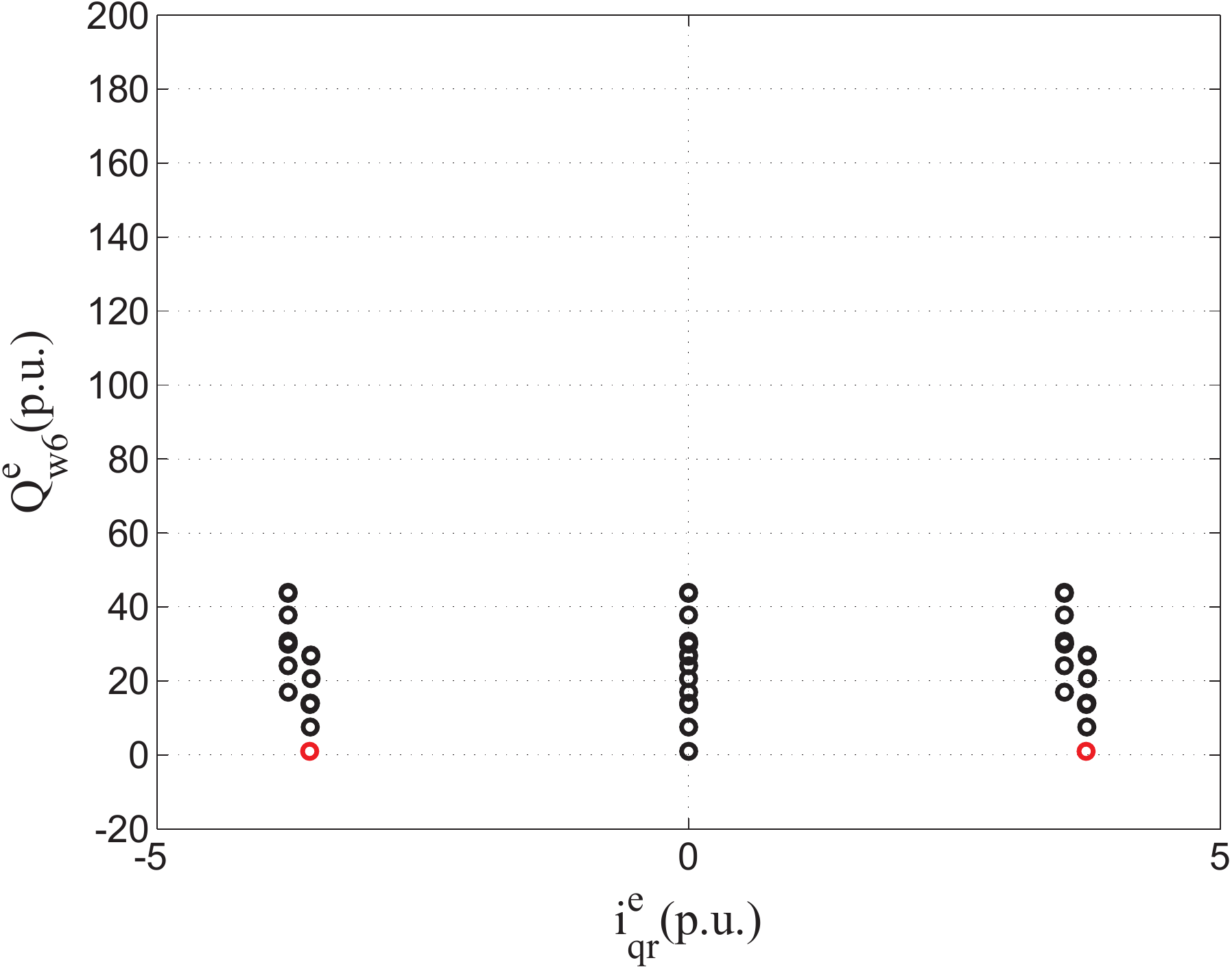}
                \caption{$\gamma$=2, $\left|V_{wj}^e\right]$=1 p.u.}
                \label{fig:mode_longwind12}
        \end{subfigure}
\caption{Equilibria for different wind penetration ($\gamma$) and wind bus voltage $\left|V_{w6}^e\right|$  }
\label{fig:main_equilibria}
\end{figure*}

A wind power plant is considered to be connected to the power system at bus 6. The plant is considered to be an aggregation of several wind turbines with their respective DFIGs connected to a single wind bus. The amount of wind power injected to the power system is controlled by the parameter $\gamma$ which when equals to 1 implies that 1000 turbines are connected to the wind bus. The parameters of the representative wind turbine are given as: $A_s=2827.43$ m$^2$, $\rho=1.225$ kg/m$^3$, $v_{r}^{e}=10$m/s, $C_{p}=0.41$, $N_{g}=78$, $w_{g}^{e}=207$ rad/s. The DFIG parameters are given as $R_s=0.0111$ p.u., $R_r=0.0108$ p.u., $L_{ls}=0.1487$p.u., $L_{lr}=0.1366$p.u., $L_m=4.6978$ p.u., $p=4$, $w_{e}=188.5$ rad/s. The active power output of the DFIG, ${P_{w6}^e}$ is assumed to be controlled by a Proportional-Integral (PI) controller to a fixed setpoint in steady-state following from the maximum power-point tracking curve. For $\gamma=1.0$, the active power output,  ${P_{w6}^e}=0.71$ p.u., when wind speed $v_{r}^{e}=10$m/s . The reactive power output, namely ${Q_{w6}^e}$, is assumed to be controlled by a different PI controller to a fixed value in steady-state so that the wind bus voltage magnitude $\left|V_{w6}^e\right|$ is maintained at a constant level. We formulate our equilibrium analysis problem by constructing \eqref{eqn:power_slbus}-\eqref{eqn:power_loadbus} using these parameter values. The solution of the equilibrium is parameterized by $\gamma$ and $\left|V_{w6}^e\right|$. The factor $\gamma$ is assumed to have values of $0.5$, $1$, $1.5$ and $2$, while $\left|V_{w6}^e\right|$ can vary among $0.96$, $0.98$ and $1$. We solve for all possible numerical roots of \eqref{eqn:power_slbus}-\eqref{eqn:power_loadbus} for different combinations of $\gamma$ and $\left|V_{wj}^e\right|$. Solution for each case specified by a certain $\gamma$ and $\left|V_{wj}^e\right|$ yields 48 different equilibria. Among these only 2 solutions have the rotor voltage $v_{qr}^e$ and $v_{dr}^e$, and the reactive power output  ${Q_{w6}^e}$ of the wind system within practical limits. In Figure \ref{fig:main_equilibria} we show all the equilibria by means of scatter plot between the rotor current $i_{qr}^e$ and ${Q_{w6}^e}$ under different scenarios.  The feasible equilibrium points are plotted in red to differentiate them from the other numerical solutions.
\quad
We next compute the small-signal model for \eqref{eqn:power_slbus}-\eqref{eqn:power_loadbus} about the two feasible equilibria for different combinations of $\gamma$ and $\left|V_{w6}^e\right|$ . The eigenvalues of the Jacobian matrix are determined to check the stability of these equilibria. Table \ref{table:stability} shows the stability property of the two feasible equilibria in different scenarios. In certain cases, one or both of the feasible equilibria are found to be unstable, justifying the use of our approach for computing all equilibria.
 Even when both the equilibria are stable the damping factors of the dominant eigenvalues change between the  different equilibrium indicating that some equilibrium to be more robust to the fluctuations in the wind penetration than others. Table \ref{table:dom_eig} shows the variation of the dominant eigenvalues for the different scenarios with two stable equilibria. The damping decreases with increasing $\gamma$ while the damping improves with higher $\left|V_{w6}^e\right|$. In summary, the results show that with increasing wind penetration - (1)  there may be multiple feasible operating points for a wind integrated
 power system, (2) the small-signal stability of the overall model around these equilibria varies
 based on the amount of wind penetration; 
  (3) increasing wind penetration may lead to undesired equilibria or loss of dynamic performance with considerably reduced damping  under certain scenarios. These observations can be very useful for power system operators in planning for control and dispatch mechanisms so that undesired power flow equilibria can be avoided under high wind penetration. 
\section{Conclusions}
\label{sec:conclu}
\begin{table*}[t]
\centering
\begin{tabular}{| p{0.1\textwidth} |p{0.2\textwidth}|p{0.2\textwidth}|}
\hline
•\centering & \centering$\gamma=0.5$&\qquad\quad\quad$\gamma=1.0$\\ 
\hline
\centering  $\left|V_{w6}^e\right|=0.96$ & \centering $\begin{aligned} &-0.07 \pm 0.71i\\ &-0.04\pm0.69i\end{aligned}$ &
 \qquad$\begin{aligned} &-0.03 \pm 0.69i\\ &-0.02\pm0.68i\end{aligned}$  \\ 
\hline 
\centering  $\left|V_{w6}^e\right|=0.98$ & \centering $\begin{aligned} &-0.09 \pm 0.71i\\ &-0.06\pm0.70i\end{aligned}$ &
 \qquad$\begin{aligned} &-0.03 \pm 0.69i\\ &-0.02\pm0.68i\end{aligned}$  \\  
\hline
\centering  $\left|V_{w6}^e\right|=1.00$ & \centering $\begin{aligned} &-0.11 \pm 0.71i\\ &-0.06\pm0.70i\end{aligned}$  & 
\qquad$\begin{aligned} &-0.06 \pm 0.70i\\ &-0.05\pm0.69i\end{aligned}$  \\ 
\hline
\end{tabular} 
\caption{Dominant Eigen values for scenarios having 2 stable equilibria}\label{table:dom_eig}
\end{table*}
In this work we have shown that the complex nonlinear model of a wind-integrated power system may exhibit multiple feasible equilibria depending on the level of wind penetration. Conventional load flow solution methods which generally estimate local solutions of nonlinear equations may fail to identify all such feasible operating points.  In our work we used a novel homotopy based continuation method to evaluate these equilibria in a computationally efficient manner. The simulation results show that with increasing wind penetration, the system may exhibit undesired or poorly damped poles, and in some instances even unstable poles. In view of the increasing renewable penetration in conventional power grid in recent times, this type of study is of immense importance as it provides system operators with all the feasible solutions of power flow equations, thereby helping them with dispatch and control. Our future work will include the application of this method to higher dimensional wind power system models with more complex dynamics such as those with nonlinear dynamic loads.


%
%
%
%
%
%
%
%
%
 
\end{document}